\documentclass[twocolumn, switch]{article} % Method A for two-column formatting

\usepackage{preprint}

\usepackage{amsmath, amsthm, amssymb, amsfonts}

\usepackage[numbers,square]{natbib}
\usepackage[utf8]{inputenc}	% allow utf-8 input
\usepackage[T1]{fontenc}	% use 8-bit T1 fonts
\usepackage{xcolor}		% colors for hyperlinks
\usepackage[colorlinks = true,
            linkcolor = purple,
            urlcolor  = blue,
            citecolor = cyan,
            anchorcolor = black]{hyperref}	% Color links to references, figures, etc.
\usepackage{booktabs} 		% professional-quality tables
\usepackage{nicefrac}		% compact symbols for 1/2, etc.
\usepackage{microtype}		% microtypography
\usepackage{lineno}		% Line numbers
\usepackage{float}			% Allows for figures within multicol

\usepackage{lipsum}		%  Filler text

\usepackage{newfloat}
\DeclareFloatingEnvironment[name={Supplementary Figure}]{suppfigure}
\usepackage{sidecap}
\sidecaptionvpos{figure}{c}

\usepackage{titlesec}
\titlespacing\section{0pt}{12pt plus 3pt minus 3pt}{1pt plus 1pt minus 1pt}
\titlespacing\subsection{0pt}{10pt plus 3pt minus 3pt}{1pt plus 1pt minus 1pt}
\titlespacing\subsubsection{0pt}{8pt plus 3pt minus 3pt}{1pt plus 1pt minus 1pt}

\title{Nanoplasmonics within a biofilm}

\usepackage{xwatermark}
\newwatermark[firstpage,color=gray!60,angle=90,scale=0.32, xpos=-4.05in,ypos=0]{\href{https://doi.org/}{\color{gray}{Publication doi}}}
\newwatermark[firstpage,color=gray!60,angle=90,scale=0.32, xpos=3.9in,ypos=0]{\href{https://doi.org/}{\color{gray}{Preprint doi}}}
\newwatermark[firstpage,color=gray!90,angle=0,scale=0.28, xpos=0in,ypos=-5in]{*correspondence: \texttt{email@institution.edu}}

\usepackage{authblk}

\author[1]{Sanhita Ray}
\author[1]{Madhurima Pattanayak}
\author[1\thanks{\tt{adgcal@gmail.com}}]{Anjan Kumar Dasgupta}

\affil[1]{Department of Biochemistry, University of Calcutta}
\begin{document}

\twocolumn[ % Method A for two-column formatting
  \begin{@twocolumnfalse} % Method A for two-column formatting
  
\maketitle

\begin{abstract}
Biofilm templated gold nanonetwork provide a platform to study transition of local plasmon to surface plasmon.  The switch from localized surface plasmon resonance (LSPR) to surface plasmon resonance (SPR) is induced by percolation of metal atoms between gold nano-islands.  Formed nano-composites showed transition in metal percolation state (non-percolating to percolating state), as evident from electrical properties. The optical component of the study followed a previously tested method by our group, in which photonic coherences are induced by biofilms. Peaks were obtained in enhancement spectra that were found to correspond with plasmonic peaks, which in turn depended on Au precursor concentration that had been added.
\end{abstract}
%\keywords{First keyword \and Second keyword \and More} % (optional)
\vspace{0.35cm}

  \end{@twocolumnfalse} % Method A for two-column formatting
] % Method A for two-column formatting

%\begin{multicols}{2} % Method B for two-column formatting (doesn't play well with line numbers), comment out if using method A

%%%%%%%%%%%%%%%  Main text   %%%%%%%%%%%%%%%
% \linenumbers

\section{Introduction}
Biofilms \cite{Donlan2002} have been used by previous workers \cite{Lengke2007} to synthesize metallic nanoforms. Nanoforms synthesized using bio-film include gold nano-spheres, triangles, silver nano-forms \cite{Lengke2007} and others. Gold nanomaterials show plasmonic activity. Plasmons are collective and coherent oscillation of electrons on the surface of gold nano-particle. On incidence of certain wavelength of light, electromagnetic energy of photon is converted to kinetic energy of plasmon \cite{Khurgin2015}. The energy conversion is maximum when frequency of incident light is same as the frequency of plasmon. This phenomenon is known as plasmon resonance. When light is incident on nano-particle the plasmons retain some energy as evanescent radiation \cite{Frumin2013}. The rest of the light is scattered due to plasmonic scattering. \\

Many workers \cite{Chen2012} haves described percolation of gold through a dielectric material. For an ensemble of gold nano-particle distribution on a dielctric surface, electron flow cannot occur when the nanoparticles are not connected. In order to achieve electron transport across the insulating materials , the nano-particles should be connected by metal atom bridges. Above a certain concentration of gold nano-partcle a sharp transition occurs showing electron transport across the matrix. This transition point is called percolation threshold. \\

Percolation is related to electrical conductivity, electric and magnetic susceptibility, optical reflectance of nanoparticle. Above percolation threshold, the localized plasmon resonances of individual islands  are coupled together in a phenomenon known as delocalization. This coherent interaction leads to changes in optical properties and light reflection. \\

Above percolation threshold, the localized plasmon resonances of individual islands are coupled together. On increasing the concentration of nanoparticles, electromagnetic energy is trapped within  plasmons and collective oscillation of plasmons is increased – this phenomenon is called surface enhancement effect which form regions of intense scattering called hotspots. By near-field optical detection method we may image such hotspot regions.\\

Biofilms show Mie scattering which is a type of elastic scattering \cite{Hahn2009} i.e. the energy of incident light is equal to the energy of the emitted light. Particles suspended within the biofilm matrix (bacterial cells) are of a diameter much greater than 10nm, and hence responsible for Mie scattering. In Mie scattering most of the incident light is scattered in forward direction and the process is energy independent.\\

Our interest is to know whether and plasmonic scattering from gold nanoparticle interacts with Mie scattering from bacterial biofilm. We have also studied whether increasing the gold nanoparticle concentration alters the extent of  light enhancement.\\

\section{Materials and Methods}
\subsection{Synthesis}
\subsubsection{Bacterial Biofilm Formation}
Rhodobacter capsulatus SB1003 (RCSB) , a non-sulphur, photosynthetic, purple bacteria , obtained as a gift from Dr. Patrick Hallenbach, was used  for formation of  bio-film which was used as a template for gold nano-particle formation from Auric Chloride.  The growth medium for bacteria used was yeast extract supplemented RCV medium (0.3g/L  yeast extract, 4g/L malic acid, 1g/L ammonium chloride, 75 mg/L calcium chloride, 1mg/L nicotinic acid, 20mg/L di-sodium EDTA, 120mg magnesium chloride, 10mM phosphate buffer at pH  6.8).\\
Transparent sheets were used as the substratum for biofilm formation. The transparent sheets were cut into small pieces and each piece was separately dipped in $70\%$ ethanol  for half- an hour for sterilization purpose and then the ethanol was dried in laminar airflow before using for biofilm growth. The sterilized sheets were then kept inside a 50ml transparent falcon which was filled with 5ml of bacterial innoculum and finally, 45ml of plain RCV medium, so that the falcon is filled upto its brim, so that bacterial growth will take place in anaerobic condition. The bacterial cells were allowed to grow for 11 days in room temperature, in the presence of artificial white light (100 watt) , keeping the falcon  horizontally , without using a shaker.\\

\subsubsection{Fixation of biofilm}
In order to fix the bacterial cells attached on the transparent sheet, the biofilm was thoroughly washed with milli Q water for several times and treated overnight  with $2.5\%$ gluteraldehyde  at 50C . Glutaraldehyde acts as a crosslinking agent. After that,  it was thoroughly washed with milli Q water to get rid of gluteraldehyde.

\subsubsection{Gold nano-particle synthesis}
Gold nano-particles were synthesized from  fixed biofilm. Different concentration of auric chloride solutions were prepared in different test tubes. Fixed biofilms were cut into small pieces. Biofilms were treated with auric chloride solution  at 50C for an hour and then  heated in 80-90 0 C temperature for 6-8 hrs. Control biofilms were immersed in milli q and subjected to same treatment. 
After exposure at 80-90C for 8 hrs, when  biofilms started tochange colour  pink to purple to violate , depending on auric chloride concentration and gold-nanoparticle formation was complete, the biofilms were washed repeatedly with milli q water to remove loosely connected nano-particles from the surface of the biofilm.

\subsection{SEM-EDAX}
For morphological characterization SEM-EDAX (Model : Zeiss EVO 18-Special Edition) were performed. Samples were either dried in an oven at 50C overnight or fixed with $2.5 \%$ glutaraldehyde, dehydrated in graded alcohol and air dried. For material characterization, no gold coating was used. Instead the substratum used was semiconducting silicon. SEM images show patches of granular nanostructures deposited on the biofilm material and their distribution mimics the cellular arrangement . Finer nanostructures were obtained when biofilm was grown on aluminium foil where nanoscale networks could be seen traversing along the biofilm cells. Atomic distributions were studied with EDAX.

\subsection{Staining with fluorophores}
The biofilms were stained with different fluorophores of different excitation and emission wavelength after synthesis of gold-nano forms. We have used different fluorophores, namely Acridine orange (Working concentration 0.001); propidium iodide ( working concentration 300 muL/ml), DAPI (). The bio-nano composites were submerged in each fluorophore for 15-30 minutes. Care was taken to avoid photobleaching. 

\subsection{Fluorescence microscopy}

Biofilms and composites were stained using 0.01 $\%$ acridine orange (AO). AO stained wells were imaged using confocal microscopy (inverted confocal microscope Olympus, FV 1000) with green filter. Z-scanning was performed and obtained images were stacked. This process ensured proper imaging of extracellular matrix. Matrix contains secreted DNA hence expected to take up AO which is a nucleic acid stain. \\

To study the effect of nanoplasmonics on fluorescence, the biofilms were stained with different fluorophores with different excitation and emission wavelength. Effect on fluorescence was studied using epifluorescence and laser scanning confocal microscopy (LSCM). \\

\subsection{Electrical characterization}
For current-voltage (I-V) and capacitance measurements RCSB biofilms were grown on gold electrodes fabricated on glass cover-slips (see Figure ~\ref{fgr:synthesis}). Biofilms grew on top of and in between the electrodes. The electrode complex with biofilm grown on top was used for gold nanoform synthesis, as previously. Current-voltage and capacitance measurements were done on Keithley system (SCS-4200 Semi-condutor Characterization System) with Cascade Microtech Probe Station gold needle electrodes. Current-voltage curves were obtained with DC voltage. Capacitance spectra were determined using AC voltage by varying its frequency. Frequency scans were run for various values of AC voltage amplitude (RMS).

\section{RESULTS}

\begin{figure*}
	\centering
	\includegraphics[scale = 0.5]{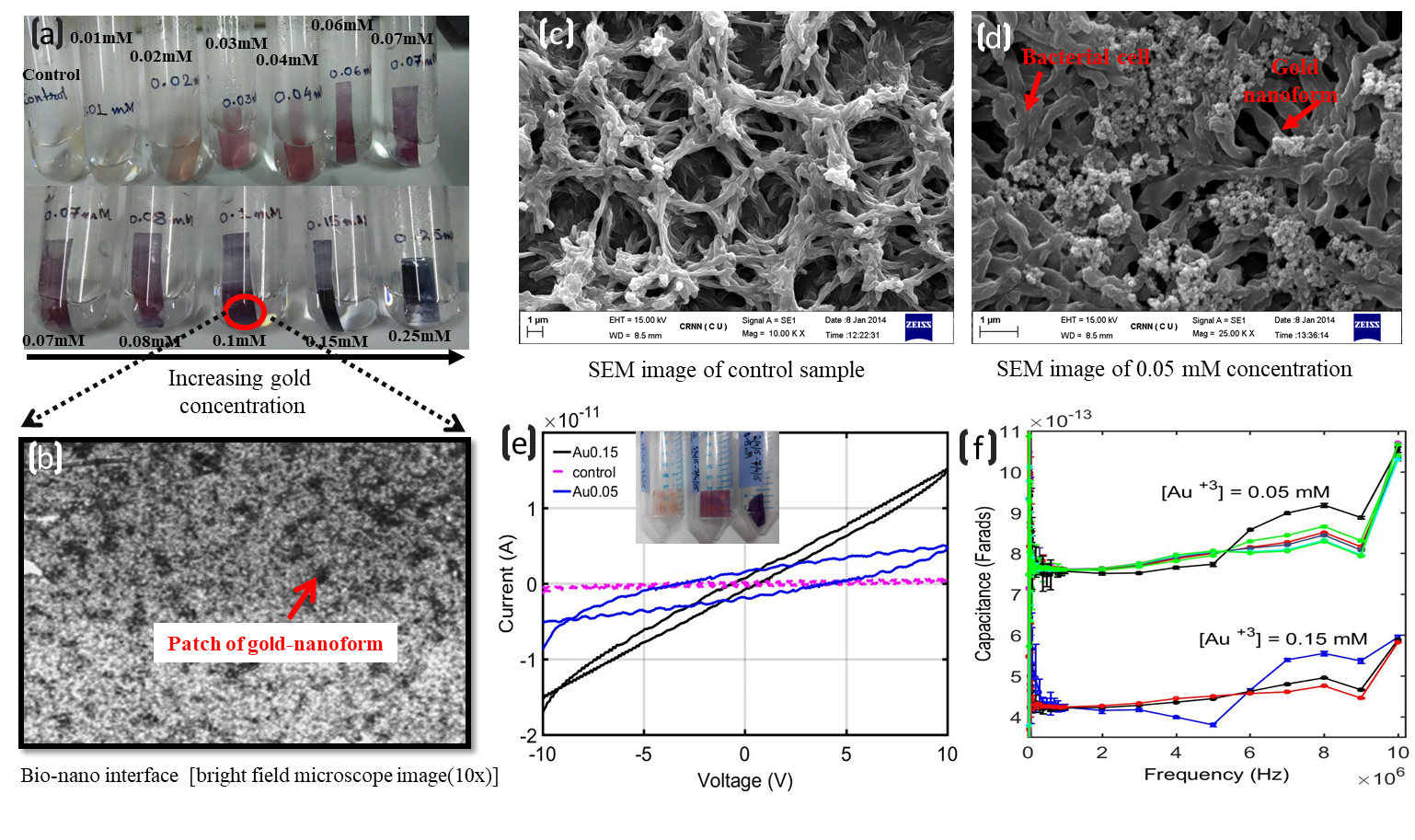}
	\caption{Figure (a) shows gold nano-structures synthesized on template of  bacterial biofilm which was grown on OHP sheet. Figure (b) shows bright field microscope image of bio-nano interface, taken in 10x magnification. SEM image (fig. c) of control biofilm. SEM image figure (d) of 0.05mM $[Au^{+3}]$; the bright patches shows presence of gold nanoform deposited on comparatively darker biofilm. The current vs voltage graph in figure (e) shows current-voltage of control, 0.15mM and 0.05 mM of $[Au^{+3}]$. Figure (f) shows capacitance vs frequency graph of 0.05 mM and 0.15 mM $[Au^{+3}]$.   }
	\label{VolFreq}
\end{figure*}

Biofilm templated gold nano-structures were formed (see figure ~\ref{VolFreq}(a-d)) upon addition of varying concentrations of gold precursor. Nano-gold was embedded within biofilm matrix and a range of pink to purple (see figure ~\ref{VolFreq}(a)) to blackish violet colored thick films were obtained. When auric chloride concentration was increased upto 0.15 mM, the color of gold-biofilm composite switched sharply from pink-purple to deep violet or almost black. Hence 0.15 mM $[Au^{+3}]$ was taken to be a threshold concentration at which the structure and plasmonic properties of formed nanomaterial changed drastically. Higher conductance and a lower capacitance of 0.15mM gold, compared to 0.05mM gold (see figure ~\ref{VolFreq}(e)) (see figure ~\ref{VolFreq}(f)) confirms the fact that 0.15mM is percolation threshold point.  \\

\subsection{Characterization}
\subsubsection{SEM imaging - EDAX}

\begin{figure*}
	\centering
	\includegraphics[scale = 0.6]{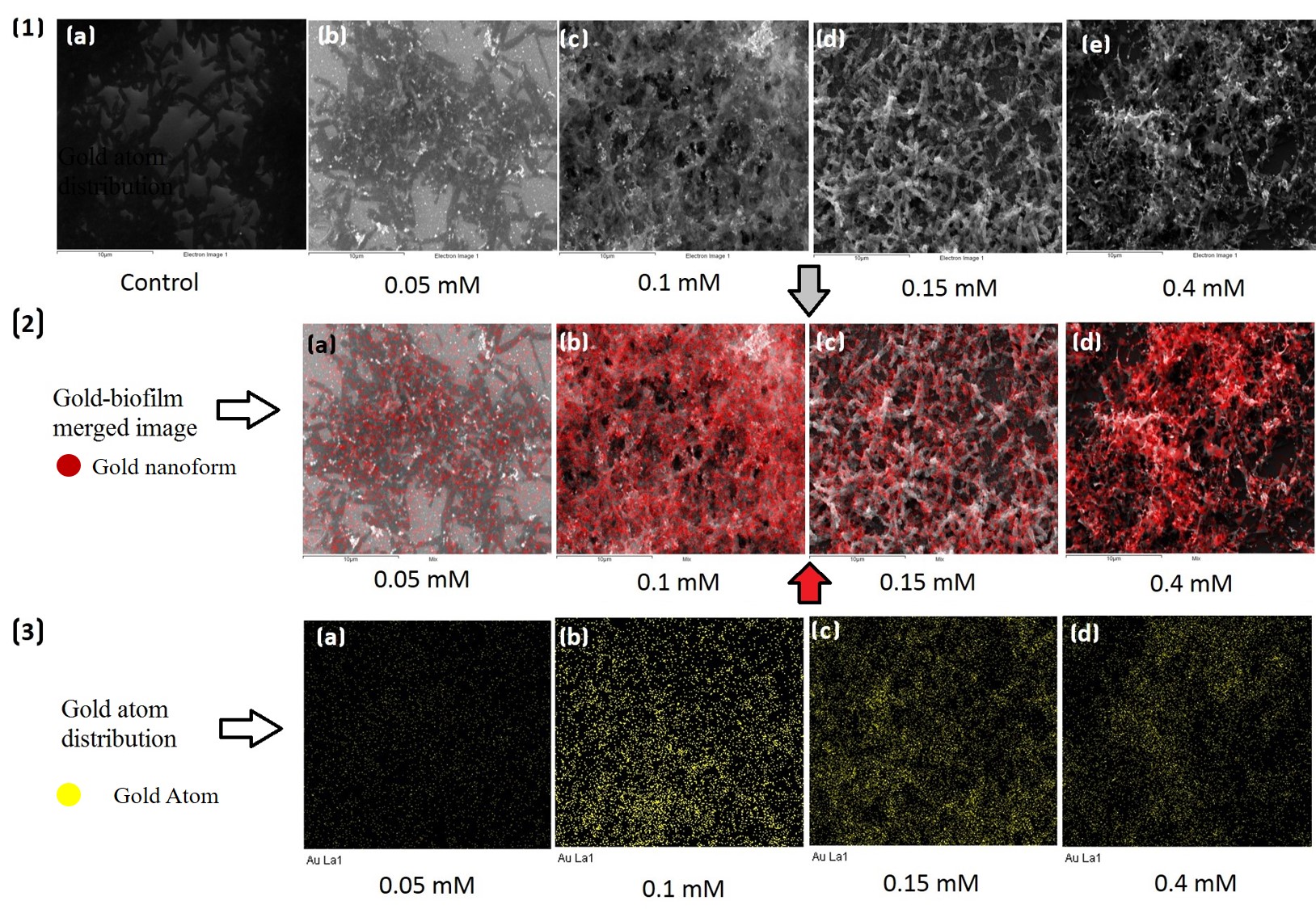}
	\caption{Figure 1(a) shows SEM image of control biofilm. Figure (1b-1c) show SEM images of biofilm with increasing gold concentration; the bright areas show presence of gold nanoforms deposited on comparatively dark biofilm template. Figure (2a-2d) obtained from EDAX mapping; gold nanoform representing red dots were merged over the respective greyscale images. Figure (3a-3d) gold atom distribution - presence of gold atom, shown in yellow colour.} 
	\label{SEM_EDX}
\end{figure*}

The detailed structure of bio-goldnano interface have been obtained using scanning electrom microscope. To obtain the structure, amount of deposition and deposition pattern of gold nanostructures, the samples were not coated with gold before performing SEM. The gold nanoforms were deposited along the biofilm template. The biofilm looks darker, since it is an insulating material (when no gold sputtering is done) and the gold nanostructures show up as bright entity against the dark background (see Figure ~\ref{SEM_EDX}(I.b.-I.e.)). Presence of gold nanostructures increased with increasing gold precursor concentration, i.e. 0.05-0.4 (see Figure ~\ref{SEM_EDX}(i.b-i.d). Above (SUPPLEMENTARY) 0.4mM concentration, amount of gold nanostructure have decreased (see Figure ~\ref{SEM_EDX}(i.e)), since the gold structures partially disintegrated due to their bigger size. \\

EDAX along with EDAX mapping was performed to determine distribution of gold atoms on the biocomposites (see figure ~\ref{SEM_EDX}(II.a.-II.d.)). Gold atom distribution of different gold precursor concentration was obtained (see figure ~\ref{SEM_EDX}(III.a.-III.d.)). The obtained SEM-EDAX images confirm the gradual formation of gold nano-network and finally percolation)

\subsubsection{Electrical properties}

Nano-biocomposites synthesized with auric chloride concentrations of 0.05 mM and 0.15 mM were grown on interdigitated electrode (see Figure ~\ref{VolFreq}(e)) (inset) .These doses were chosen for electrical characterization. Sharp change in color was shown at 0.15 mM., gold incorporated biofilms showed higher current conduction (see Figure ~\ref{VolFreq}(e)) than control biofilm. This is expected since any available gold patch would serve as a short circuit through the dielectric material (biofilm matrix). For the lower precursor concentration, extent of hysteresis (area within the curve, when voltage is looped) was greater compared to that for higher concentration of gold. This indicates higher charge storing capacity of gold-biofilm composite when gold concentration is below the threshold. When gold concentration exceeded threshold, conductive pathways were formed within the insulating biofilm, thus dissipating any stored charge.\\

Capacitance measurements (see Figure ~\ref{VolFreq}(f)) with AC voltage showed presence of convergence points (isosbestic points) in gold-biofilm composites. Capacitance values increased when biofilms were treated with 0.05 mM auric chloride, compared to control films. However, for higher dose (0.15 mM) of $[Au^{+3}]$, capacitance values decreased, compared to lower dose (0.05 mM $[Au^{+3}]$). This supports the conclusion that at 0.15 mM $[Au^{+3}]$, metal percolation occured through the biofilm, providing an alternate pathway for electron flow through the material (parallel resistor, as opposed to resistor in series with capacitor).\\

\subsection{Light enhancement studies}

\begin{figure*}
	\centering
	\includegraphics[scale=0.5]{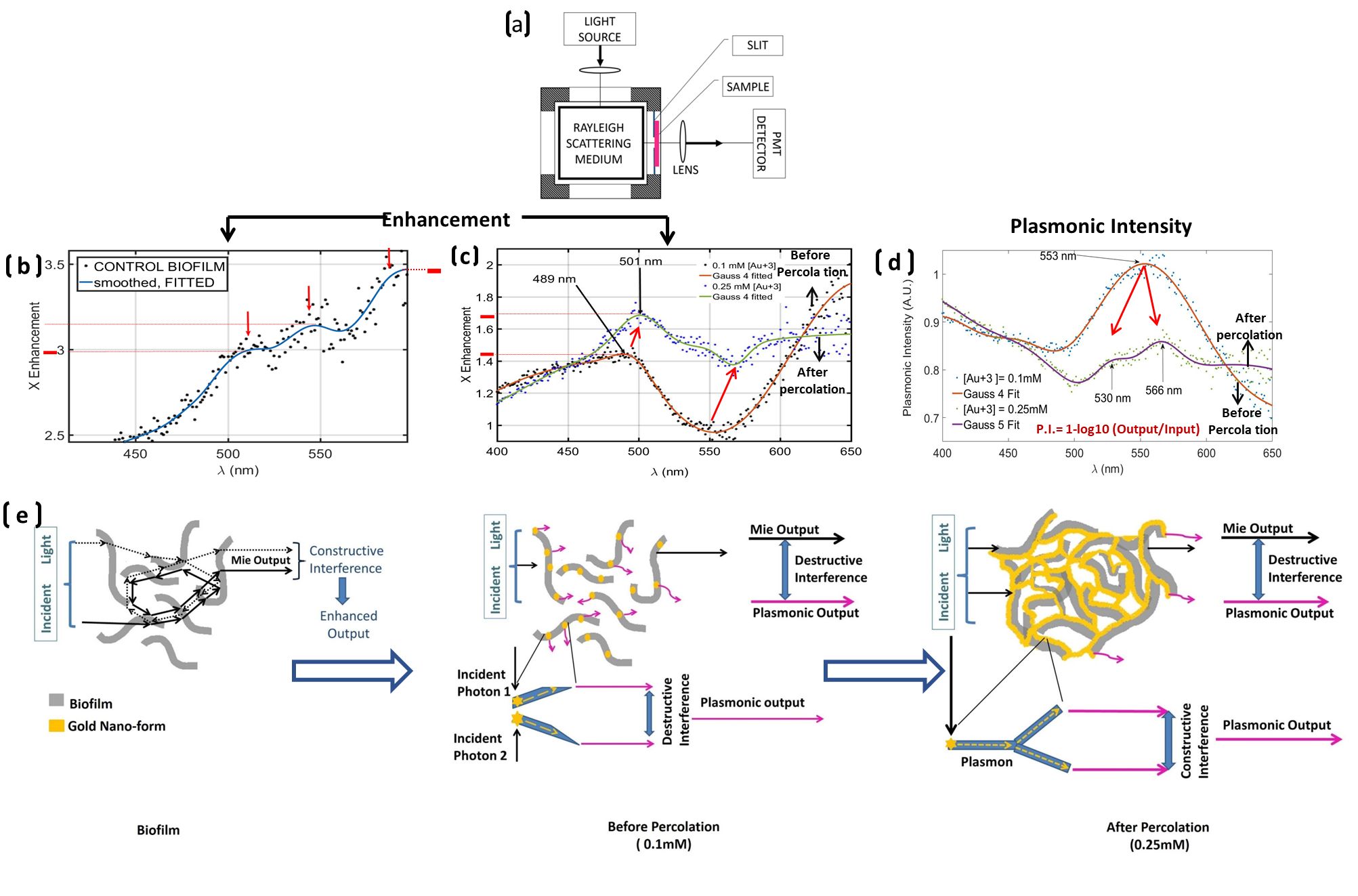}
	\caption{Figure (a) showing diagram and mechanism of PTI. Figure (b) showing enhancement of controlled biofilm - the red arrow show the enhancement peaks. Figure (c) showing enhancement vs wavelength graph of 0.1mM (before percolation) and 0.25mM (after percolation) of auric chloride. Figure (d) shows plasmonic vs wavelength graph- plasmonic intensity of 0.1 mM (before percolation) and 0.25mM (after percolation) of Auric chloride have been shown. Fig(e.i) shows type of light scattering from control biofilm. Figure (e.ii) shows destructive interfarence between Mie Scattering and plasmonic scattering. Figure (e.iii) shows destructive interfarence between mie scattering and plasmonic scattering after percolation. }
	\label{Plasmon}
\end{figure*}

Our previous findings show that when light is passed (see Figure ~\ref{Plasmon}(a)) through Rhodobacter biofilm, an amplified output is obtained. When times enhancement (output/input) is plotted as a function of wavelength, we refer to the resultant graph as an enhancement spectrum (see Figure ~\ref{Plasmon}(b-c)). In this paper we aimed to find out the resultant output when plasmonic nanoparticles were embedded within the biofilm matrix. When gold-biofilm composites (see Figure ~\ref{Plasmon}(c)) were placed in the path of Rayleigh scattered light, enhanced output was obtained but the extent of enhancement was less than control biofilm in all cases.\\

For specific concentration range (from 0.05 mM to 0.4 mM auric chloride), enhancement spectra showed peaks (see Figure ~\ref{Plasmon}(c)) corresponding to regions where there is minimal plasmonic scattering. Enhancement peaks were present at around 500 nm and 650 nm. Troughs were present  where resonant plasmonic scattering occured. Plasmonic scattering profiles (see Figure ~\ref{Plasmon}(d)) could be obtained by taking the log of enhancement and substracting it from 1. For concentration range, from 0.05 mM to 0.4 mM auric chloride, a broad plasmonic peak was observed which could be resolved by Gaussian fitting into 3 or 4 component peaks.\\ 

\begin{figure*}
	\centering
	\includegraphics[scale=0.2]{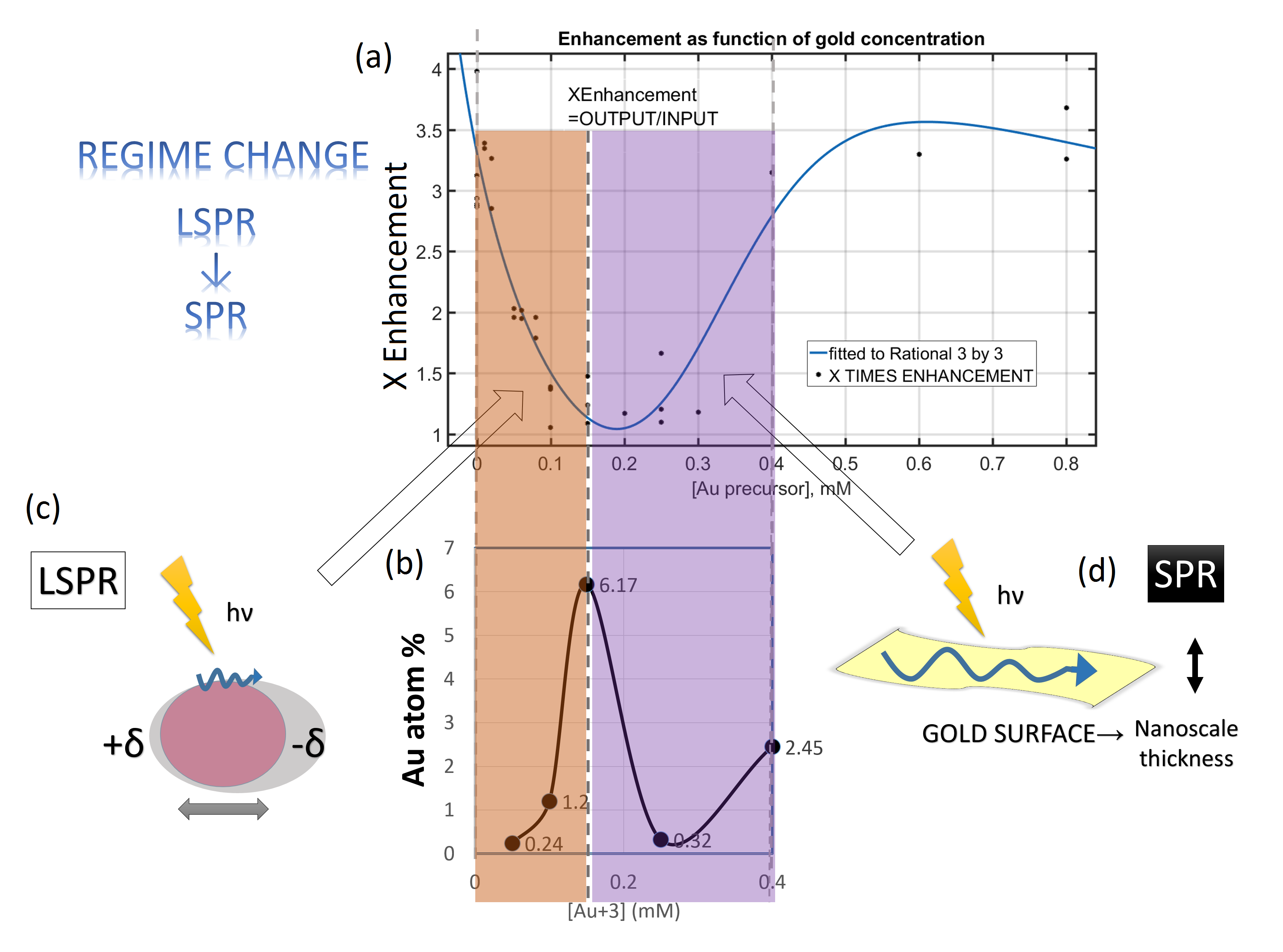}
	\caption{(a) Times enhancement as a function of gold precursor concentration added. (b) Gold atom percent as measured by EDAX as a function of gold precursor concentration added. (c-d) Schematic showing LSPR to SPR transition corresponding to regime change in enhancement curve}
	\label{fig_four}
\end{figure*}

Peak heights at 500 nm were plotted against auric chloride concentrations used (see Figure ~\ref{fig_four} (a)). Thus enhancement was found to be critically dependent on gold precursor concentration. Below the threshold concentration i.e. below 0.15 mM $[Au^{+3}]$, extent of enhancement decreased when concentration was increased. Above the threshold, the degree of enhancement increased with increasing concentration. The concentration dependence was found to fit a rational function. Rational functions have been previously used to simulate threshold circuits.\\

This enhancement function is highly sensitive to variations in the biofilm template. Enhancement pattern was strikingly different when the biofilm was grown for lesser number of days (7 as opposed to 11 days of growth, see supplementary Figure 2). For higher concentrations, where no peaks were present, substracting control spectrum from sample spectrum yielded a difference spectrum that showed peaks corresponding to resonant plasmonic peaks. 
%This further proves that Mie scattering (strongest at 0 degree) is responsible for enhancement.

\subsection*{Fluorescence microscopy}

When observed under a fluorescence microscope (with non-coherent lamp illumination), acridine orange (AO) stained nanogold-biofilm composites showed a blurred green fluorescence (see supplementary Figure 3) through out the sample, even though focussing was carefully adjusted. For higher precursor concentrations, some indistinct structures were identifiable. In bright field mode, the same samples showed presence of distinct cells. Control samples, which had undergone identical heat treatment as the samples, showed the presence of distinct cells when observed in both bright field and fluorescence modes. Therefore we may conclude that cellular stuctures have not undergone disintegration in case of composites. Rather the near field scattering within gold biocomposites have scrambled up the expected fluorescent microscopic image. Strong, coupled near field scattering is associated with development of plasmonic hotspots.\\

\begin{figure*}
	\centering
	\includegraphics[scale=0.9]{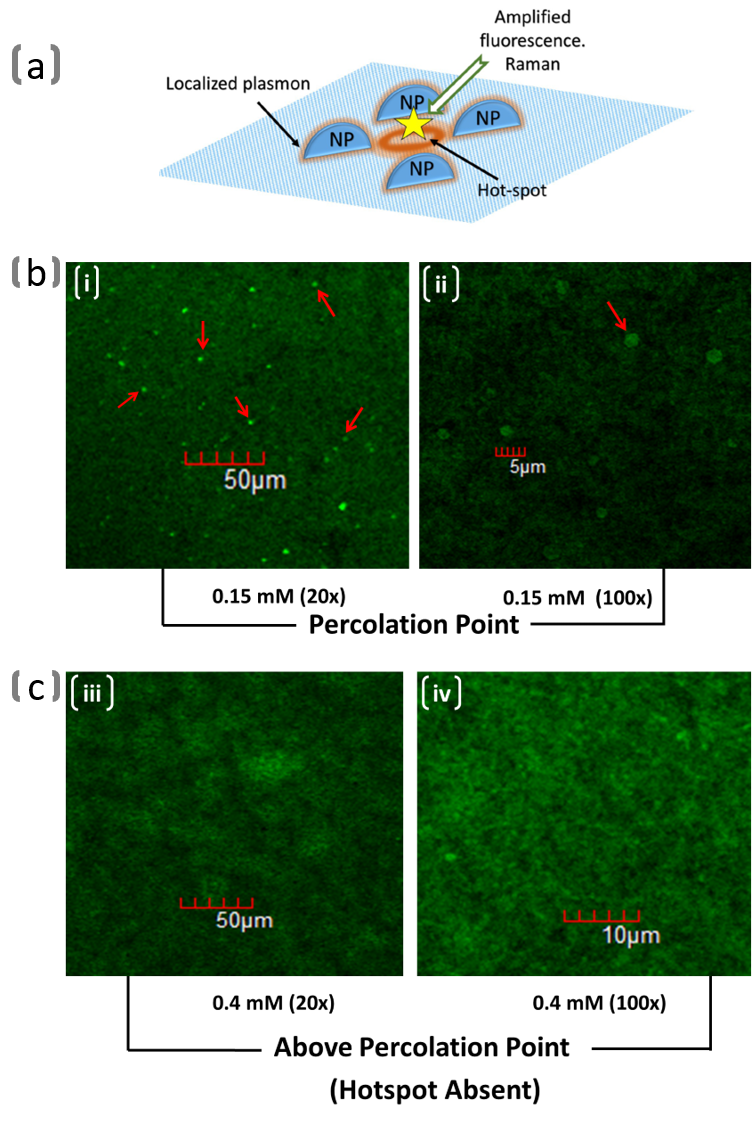}
	\caption{Schematic diagram (Figure a) shows localized plasmon phenomenon and hotspot formation. Presence of hotspot in 20x (Figure B.i.) and 100x (Figure B.ii.) under laser scanning confocal microscope (LSCM). Figure c showing presence of delocalized surface plasmon above percolation threshold in 20x (Figure c.i.) and 100x (Figure c.ii) magnification under LSCM.}
	\label{fgr:confo}
\end{figure*}

Confocal microscopy (with coherent laser illumination) revealed the presence of distinct hot-spots (see Figure ~\ref{fgr:confo}(a)) for $Au^{+3}$ = 0.15 mM. The circular hotspots (marked with orange arrows, see figure ~\ref{fgr:confo}(b-c)) may be seen clearly. When 3D reconstruction of the z-sections were done, it showed hotspots as a column of uniform, high fluorescence intensity. When confocal images were obtained for higher precursor concentration ($Au^{+3}$ = 0.4 mM), instead of seperate hotspots, high fluorescence intensity was observed all over the sample as a patchy network (~\ref{fgr:confo}(d-e)). We propose that the percolated gold nanonetwork acts as an optical trap. Thus the AO fluorescence undergoes multiple cycles of reflection before being released from the biocomposite. Multiple reflections result in amplification of resultant light intensity.\\

The different results with fluorescence microscope using a lamp illumination (non-coherent) versus a confocal laser scanning microscope (coherent illumination) suggest that coherence properties of incident radiation is an important factor determining photon transport through the material.\\

\section{Discussions}

We have used glutaraldehyde fixed bacterial biofilm to synthesize gold nanoparticle from auric chloride. Glutaraldehyde is a crosslinking agent that insolubilize the cellular proteins and fixes the cell. Hence, glutaraldehyde fixes free proteins and poly-saccharides in the biofilm. Polysaccharide is the main component that form the matrix of the biofilm and keeps the bacterial cells intact within the matrix. The polysaccharides, proteins of bacterial cells act as reducing agents which reduce auric chloride into gold nano-particles (GNP). In the unfixed biofilm, the proteins and amino acids do not insolubilize and fix, due to which free amino acids, as well as free poly saccharides remain in the biofilm. The free poly-saccharides and amino acids form a coat on GNP. \\
We have observed that the colour of synthesized biocomposite varied upon the concentration of auric chloride and the colour varied from purple to violet to dark violet. Upon increasing the concentration of auric chloride, gold atom incorporation into biofilms increased but decreased after a critical precursor concentration. After being reduced AuCl3 forms gold nanoislands (low concentrations) and then gold nanonetworks (GNN). SEM and EDAX images show that at 0.02mM concentration of AuCl3, GNPs have started to deposit close to each other. When the concentration of AuCl3  was as high as 0.1mM, GNPs were linked to form gold nano networks. \\

450-650 nm of light was incident on biofilm-GNN composite to study interaction between Mie scattering from biofilm and plasmonic scattering of GNN. The bacterial cells being many folds larger than 10nm, show Mie scattering when incident light falls on it \cite{Hahn2009}. 450-650 nm of light was incident on biofilm-GNN composite to study interaction between Mie scattering from biofilm and plasmonic scattering of GNN. Upon gold incorporation, light amplification by biofilm decreased. At 0.15mM of auric chloride concentration a sharp transition occurs, showing an increase in scattering. This transition point is called percolation threshold. Above the percolation threshold the magnitude of scattering has increased on gradual increment of auric chloride concentration. \\

At percolation threshold i.e. where biofilm treated with 0.15mM of auric chloride, the localized plasmon resonances of individual islands are coupled together. Due to close arrangement of nanoparticles, more electromagnetic energy is trapped within  plasmons and collective oscillation of plasmons is increased and as such forms a hotspot which have been observed by confocal fluorescence microscopy \cite{doi:10.1021/jp209605v}. \\

When concentration of precursor was 0.4mM in fixed biofilm, we observed that the whole surface of biofilm was covered with GNN. Therefore, no localization effect of plasmon was present. Instead, the whole assemblage of GNN on the biofilm oscillated as single unit, showing delocalization phenomenon. Above 0.6 mM AuCl3 concentration, GNP deposition was so high that the newly formed GNPs clustered together forming larger clump of nanoparticles and could no longer remain attached to the surface of biofilm; due to which biofilms treated with more than 0.06 mM AuCl3 show lesser amount of gold nanoparticles. \\

Hotspots were found to increase fluorescence amplification. Any fluorescent phenomenon linked to analyte detection may be observed at very low analyte concentrations, if the whole machinary is present within the biofilm.\\

In this work we have shown transition of unconnected gold nanopatches to connected assemblies of Au nanonetworks, due to percolation of gold atoms on the biofilm template. Biofilm formation itself is regarded as percolation of bacterial cells on a given substratum, thus the template provides a connected background on which to place gold atoms. This process may be simulated by using a lattice model where the initial matrix is a pre-formed connected network.\\

The direct proof of percolation transition lies in studying the distribution of Au atoms on biofilm template by SEM-EDAX mapping. The variation of Au atoms concentration undergoes a sharp inflection point at an auric chloride (gold precursor) concentration which roughly corresponds to similar inflection point in optical behavior.  \\

There is a clearly marked regime change in the optical behavior of the gold-biofilm nanohybrids. The switch is from confined plasmonic fields to a delocalized plasmon or plasmonic waveguide.\\

The photosynthetic biofilm template itself induces coherence among output photons and as a result the output irradiation is greater than input by upto 4 times. A previous publication describes this phenomenon in detail. The enhancement is due to constructive interference (C.I.) among multiply scattered photons emerging from the biofilm. C.I. can occur only when there is phase coherence among output photons. \\

Plasmonic scattering occurs at a phase shift of $\pi$ with respect to incident irradiation. Since input phase is randomized by Rayleigh scattering (by water in the cuvette) the probability of destructive interference with output photons is higher. Hence enhancement is attenuated when gold atom concentration is increased.\\ 

At very high gold concentrations, bigger gold structures can no longer be stable and dissociate from the film in the form of a precipitate. EDAX results show that in such conditions Au atom percentage in the nanohybrids decreases compared to synthesis with lesser precursor concentration. This may be one factor why enhancement is less attenuated at higher gold concentrations.\\

Other than this there was a significant change in the plasmonic profiles after percolation. There was emergence of multiple peaks demonstrating rod like entities of various lengths as well as red shift of peaks. This signifies a collection of elongated structures instead of particles.\\

Electrical properties show a supporting regime change from increasing capacitance to a regime where capacitance decreases with increasing Au atom deposition.\\

\section{CONCLUSION}

Biofilm templated gold nanomaterials showed a percolation like transition in the optical plane, when isolated nanoislands were connected into nanonetworks. Optical enhancement through biofilm corresponded to such critical transition. Electrical measurements confirm a similar critical behavior.

%%%%%%%%%%%% Supplementary Methods %%%%%%%%%%%%
%\footnotesize
%\section*{Methods}

%%%%%%%%%%%%% Acknowledgements %%%%%%%%%%%%%
%\footnotesize
%\section*{Acknowledgements}
\section*{Conflicts of interest}
There are no conflicts to declare.

\section*{Acknowledgements}
The authors would like to acknowledge confocal microscopy facility of DBT-IPLS, and SEM facility at CRNN, University of Calcutta. This work has been presented as a poster at E-MRS Spring Meeting 2018 and won the Young Scientist Award for the symposium.

%%%%%%%%%%%%%%   Bibliography   %%%%%%%%%%%%%%

%\bibliography{all_plasmon} %You need to replace "rsc" on this line with the name of your .bib file
%\bibliographystyle{rsc} %the RSC's .bst file
%\bibliographystyle{plain}
%%%%%%%%%%%%  Supplementary Figures  %%%%%%%%%%%%
%\clearpage

%%%%%%%%%%%%%%%%   End   %%%%%%%%%%%%%%%%
%\end{multicols}  % Method B for two-column formatting (doesn't play well with line numbers), comment out if using method A
\end{document}